\documentclass[preprint,aps]{revtex4}
\usepackage{graphicx}
\newcommand{\be}{\begin{equation}}
\newcommand{\ee}{\end{equation}}
\newcommand{\ba}{\begin{eqnarray}}
\newcommand{\ea}{\end{eqnarray}}
\newcommand{\p}{\partial}
\newcommand{\f}{\frac}

\begin{document}
\title
{Analysis of light-wave nonstaticity in the coherent state
 \vspace{0.0cm}}
\author{Jeong Ryeol Choi\footnote{E-mail: choiardor@hanmail.net } \vspace{0.0cm}}

\affiliation{Department of Physics, Kyonggi University, 
Yeongtong-gu, Suwon,
Gyeonggi-do 16227, Republic of Korea \vspace{0.0cm}}

\begin{abstract}
The characteristics of nonstatic quantum light waves
in the coherent state in a static environment is investigated.
It is shown that the shape of the wave varies periodically
as a manifestation of its peculiar properties of nonstaticity like the case of the Fock-state
analysis for a nonstatic wave.
A belly occurs in the graphic of wave evolution whenever the wave is maximally displaced
in the quadrature space, whereas a node takes place every time
the wave passes the equilibrium point during its oscillation.
In this way, a belly and a node appear in turn successively.
Whereas this change of wave profile is accompanied by
the periodic variation of electric and magnetic energies, the total energy is conserved.
The fluctuations of quadratures also vary in a regular manner according to the wave transformation in time.
While the resultant time-varying uncertainty product is always larger than (or, at least, equal to)
its quantum-mechanically allowed minimal value ($\hbar/2$),
it is smallest whenever the wave constitutes a belly or a node.
The mechanism underlying the abnormal features of nonstatic light waves
demonstrated here can be interpreted by the rotation of the squeezed-shape contour of the Wigner
distribution function in phase space.
\vspace{0.5cm} \\
\end{abstract}

\maketitle

{\ \ \ } \\
{\bf INTRODUCTION
\vspace{0.2cm}}
\\
A light wave in media can vary
spatiotemporally according not only to the variation of electromagnetic parameters
but to its interaction with matters as well.
This may lead the wave being nonstatic \cite{lib,797,4422,fop,ldw2,usl,mno,kal}.
Regarding this, the interaction of laser light with time-varying
media has been a subject of great interest
from the early days of modern physics \cite{gzs,gzs2,fop,ldw1,ldw2,kal,ufr}.
Electromagnetic waves can be amplified or dissipated through the coupling of them,
for example, with a plasma wave \cite{fop,gzs2,ldw2,kal,ufr}.
This outcome is applicable to several physical branches,
such as the frequency shifts of light waves \cite{kal,ufr}, laser-driven wakefield accelerators \cite{ldw1,ldw2},
plasma parametric amplification \cite{usl,1012}, and harmonic generation \cite{hg}.
A notable field among them is a production of terahertz/millimeter-waves via frequency shifts.
The resources of other means for  producing such waves are actually rare and limited \cite{mno}.
It is also noteworthy that femtosecond light pulses produced by high power lasers
are necessary in realizing ultrafast optical technology
such as ion acceleration at multi-MeV energies \cite{iac1,iac2}.
Beside these, there are many other scientific and technological branches
where the nonstatic waves are utilizable \cite{npo1,npo3,npo6,1608}.

From our recent report associated with light-wave nonstaticity \cite{nwh},
it was known 
that nonstatic waves also take place even when the environment is static.
Regarding this, the time behavior of nonstatic waves in the
Fock states was analyzed fundamentally.
Nonstatic waves in such a case show a peculiar property that the waves undergo collapse
and expansion in turn periodically in quadrature space.
The concept of the measure of nonstaticity has been introduced as
a tool for estimating the magnitude of nonstatic character for such a wave \cite{nwh}.
In order for efficient manipulation and control of nonstatic waves, it is necessary to understand the
mechanism how nonstatic light waves evolve.

The research for wave nonstaticity in a coherent state, as well as in the Fock states,
may also deserve our attention.
Coherent states are fundamental in quantum optics because they allow classical-like description
of light waves.
Glauber-type coherent states \cite{qtc} provide an elegant representation of wave evolution
with Gaussianity.
A paradigmatic research with coherent states is demonstrating quantum-classical
correspondence which addresses how quantum behavior of a system develops to classicality \cite{qcc}.
It is well known that coherent-state description of a quantum state
can also be extended to a wide range of branches in physics
beyond quantum optics, such as atomic physics \cite{acs,acs2}, nuclear physics \cite{nc1,nc2},
solid state mechanics \cite{sc0,sc1}, biological systems \cite{bc2,bc3}, etc.

Stimulated by the above consequence and associated
research trends, we will investigate nonstatic waves in this work,
focusing on the coherent state in a static environment.
A quadratic invariant operator which follows the Liouville-von Neumann equation
will be adopted for this purpose.
Lots of dynamical properties of non-ideal physical systems including nonstatic light waves
can be treated by means of such a dynamical invariant \cite{le2,pos,tqd}.
The reason why the invariant operator method is useful in this context is that a generalized
quantum wave function of a light wave is obtained by utilizing
an invariant operator instead of the direct use of the Hamiltonian.

How the wave nonstaticity affects the time evolutions of diverse physical quantities
will be analyzed rigorously.
The similarities and differences between the evolutions of nonstatic and static coherent waves
will be clarified.
Moreover, we will find a global profile of wave nonstaticity extended to the coherent state
and provide graphics of the wave evolutions which display apparent nonstaticity.
The phenomena of wave collapse and expansion in the coherent state will be illustrated
and compared with those occurred in the Fock states.
The behavior of quantum energy and quadrature fluctuations which accompany nonstatic coherent waves
will also be addressed, examining their pattern/regularity in the evolution.
Finally, we will try to elucidate the mechanism related to resultant
wave-nonstaticity on the basis of the Wigner distribution function.
\\
\\
\\
\\
{\bf RESULTS AND DISCUSSION  \vspace{0.2cm}} \\
{\bf Basic Formulation and the Invariant Operator} \\
We consider a light wave propagating through a non-conductive medium in which
electric permittivity $\epsilon$ and magnetic permeability $\mu$ do not vary over time.
Because the electromagnetic parameters are independent of time in this case,
the medium is static to the wave packets evolving in it.
The Hamiltonian for the light waves in such a static medium is given by
$
\hat{H}={\hat{p}^2}/{(2\epsilon)} + \epsilon\omega^2 \hat{q}^2/2 ,
$
where $\omega$ is the angular frequency of the form $\omega =k c$ whereas $k$
is the wave number.
Then, the wave velocity is constant and it is given by $c=1/\sqrt{\epsilon\mu}$.

By the way, as mentioned in the introduction part, nonstatic quantum waves can
emerge in the Fock states in this static situation.
Wave nonstaticity may also appear in other states in general,
such as the coherent state, the squeezed state, and the thermal state.
We investigate quantum wave phenomena associated with the
nonstatic coherent state built up under
the static circumstance. This will be carried out based on the complete
analytical description of them.

To treat the light in a general way, let us see the invariant operator theory \cite{le2}.
In fact, the Hamiltonian itself is an invariant operator for this system, because the
Hamiltonian is a time-{\it in}dependent form that corresponds to a conserved energy.
However, in order to treat the system more generally,
we need to obtain a general form of an invariant $\hat{I}$ from the Liouville-von Neumann
equation:
\be
{d \hat{I}}/{d t} = {\p \hat{I}}/{\p t} +  [\hat{I},\hat{H}]/(i\hbar) = 0.
\ee
By inserting the Hamiltonian into this equation, we derive a quadratic invariant operator as
\be
\hat{I}= \hbar\omega\left[ \f{\epsilon \omega}{2\hbar f(t)} \hat{q}^2
+\f{f(t)}{2\epsilon \omega \hbar} \Bigg( \hat{p}-\f{\epsilon \dot{f}(t)}{2f(t)}\hat{q}\Bigg)^2 \right],
 \label{2}
\ee
where $f(t)$ is a time function that yields the nonlinear equation \cite{pos,nwh}
\be
\ddot{f}(t) - \f{[\dot{f}(t)]^2}{2f(t)} + 2\omega^2 \left(f(t)- \f{1}{f(t)}\right) =0. \label{3}
\ee
Although the time derivative of Eq. (\ref{2}) results in zero, the invariant operator $\hat{I}$
is given in terms of $f(t)$ that is related to the time evolution of the system.
We consider a general solution for $f(t)$, which is given by \cite{nwh}:
\be
f(t) = c_1 \sin^2 \tilde{\varphi}(t)+ c_2 \cos^2 \tilde{\varphi}(t) +
c_3 \sin [2\tilde{\varphi}(t)], \label{4}
\ee
where $\tilde{\varphi}(t)=\omega (t-t_0) +\varphi$, $t_0$ is an initial time and $\varphi$ is a phase,
while $c_1$, $c_2$, and $c_3$ are real constants that follow the condition
\be
c_1c_2-c_3^2 = 1,~~~~~~c_1c_2 \geq 1. \label{5}
\ee
Without loss of generality, we restrict $\varphi$ within the range
$-\pi/2 \leq \varphi < \pi/2$  for convenience;
the consideration of this range is enough because the period of Eq. (\ref{4}) is $\pi$.

At this stage, we introduce an annihilation operator associated with
the invariant, Eq. (\ref{2}), which is of the form \cite{pos}
\be
\hat{A}= \sqrt{\f{\epsilon\omega}{2\hbar f(t)}}\left(1- i \f{\dot{f}(t)}{2\omega}\right) \hat{q} + i\sqrt{\f{
f(t)}{2\epsilon\omega\hbar}} \hat{p}.  \label{7}
\ee
Then, its Hermitian adjoint $\hat{A}^\dagger$ is a creation operator.
These operators obey the boson commutation relation $[\hat{A},\hat{A}^\dagger]=1$.
Notice that the invariant operator can be rewritten in terms of these generalized ladder operators to be
\be
\hat{I}= \hbar\omega\bigg( \hat{A}^\dagger \hat{A} +\f{1}{2}\bigg). \label{6}
\ee
If we denote the eigenfunctions of $\hat{I}$ as $\langle q |\Phi_n(t) \rangle$ ($n=0,1,2,\cdots$),
we can obtain them from the eigenvalue equation
$\hat{I}\langle q |\Phi_n(t) \rangle = \lambda_n \langle q |\Phi_n(t) \rangle$,
while their formula is provided in Eq. (\ref{M33}) in {\bf METHODS} section (the last section).
Here, the eigenvalues are given by $\lambda_n = \hbar\omega ( n +{1}/{2})$.
Notice that the wave functions in the Fock states are represented by $\langle q |\Phi_n(t) \rangle$
as shown in Eq. (\ref{M32}).
For the basic of solving the eigenvalue equation of an invariant operator, refer to Refs. \cite{le1,le2}.

From inverse representations of Eq. (\ref{7}) together with its conjugate equation ($\hat{A}^\dagger$),
we can readily have the formula of $\hat{q}$ and $\hat{p}$, such that
\ba
\hat{q} &=& \sqrt{\f{\hbar f(t)}{2\epsilon\omega}}
(\hat{A} + \hat{A}^\dagger),  \label{6-1}
\\
\hat{p} &=& -i \sqrt{\f{\hbar \epsilon
\omega}{2f(t)}} \left[\left(1+i
\f{\dot{f}(t)}{2\omega}\right)\hat{A} -\left(1-i
\f{\dot{f}(t)}{2\omega}\right) \hat{A}^\dagger \right]. \label{7-1}
\ea
These formulae are useful when we investigate the behavior of quantum observables
such as quadrature fluctuations and quantum energy.

As is well known, the Glauber coherent state is the eigenstate of an annihilation operator.
A generalized wave function in the coherent state will be derived
by evaluating the eigenstate of $\hat{A}$ given in Eq. (\ref{7}) in the subsequent subsection.
We will use this wave function as a basic tool for unfolding quantum theory of wave nonstaticity.
\\
\\
{\bf Wave Nonstaticity in the Coherent State} \\
To obtain the analytical description of the wave function in the coherent state,
let us see the eigenstate of $\hat{A}$.
If we write the eigenvalue equation of $\hat{A}$ in the form
\be
\hat{A} |A \rangle =A |A \rangle, \label{8}
\ee
$|A \rangle$ is the coherent state.
By solving Eq. (\ref{8}) using Eq. (\ref{7}) in the configuration space in a straightforward way,
we have the coherent state as
\be
\langle q |A \rangle = \sqrt[4]{\f{\zeta(t)}{\pi }}
\exp \Bigg[ -\f{\zeta(t)}{2} \bigg( 1-
i \f{\dot{f}(t)}{2\omega} \bigg)q^2 + \sqrt{{2\zeta(t)}} A q
-\f{1}{2} |A|^2-\f{1}{2} A^2 \Bigg], \label{19}
\ee
where $\zeta(t) = {\epsilon\omega}/{[\hbar f(t)]}$.
From this wave function, we can investigate various properties of the nonstatic light wave
in the coherent state.

If $c_1=c_2=1$ and $c_3=0$, the wave undergoes no nonstaticity.
The wave nonstaticity occurs only when $c_1$ and/or $c_2$ deviate from unity.
If such deviations are large, the wave becomes highly nonstatic.
Regarding this, the measure of nonstaticity associated
with Eq. (\ref{19}) is the same as that in the Fock states, which is of the form \cite{nwh}
\be
D_{\rm } = \f{\sqrt{(c_1+c_2)^2-4}}{2\sqrt{2}}.  \label{}
\ee

We now see the eigenvalue $A$ of Eq. (\ref{8}) in detail.
For this purpose, let us denote the solutions of the classical equations of motion
(second-order differential equations) for canonical variables $q$ and $p$
as $Q_{\rm cl}(t)$ and $P_{\rm cl}(t)$, respectively.
Then, the eigenvalue is given in terms of them as
\be
A= \sqrt{\f{\epsilon\omega}{2\hbar f(t)}}\left(1- i \f{\dot{f}(t)}{2\omega}\right) Q_{\rm cl}(t) + i\sqrt{\f{
f(t)}{2\epsilon\omega\hbar}} P_{\rm cl}(t).  \label{9}
\ee
From
fundamental mechanics, we can represent
\ba
Q_{\rm cl}(t) &=& Q_0 \cos\tilde{\theta}(t),  \label{10}  \\
P_{\rm cl}(t) &=& \epsilon \f{d Q_{\rm cl}(t)}{dt}, \label{11}
\ea
where $\tilde{\theta}(t)=\omega (t-t_0) +\theta_0$, $\theta_0$ is an arbitrary phase at $t_0$.
The eigenvalue $A$ can be rewritten in terms of an amplitude and a phase, such that
\be
A=A_0 e^{i\kappa}, \label{12}
\ee
where
\ba
A_0 &=& \Bigg\{ \f{\epsilon\omega}{2\hbar}\Bigg[\f{\cos^2 \tilde{\theta}(t)}{f(t)}+
\Bigg( \f{\dot{f}(t)}{2\omega\sqrt{f(t)}}\cos \tilde{\theta}(t)
+\sqrt{f(t)}\sin \tilde{\theta}(t) \Bigg)^2 \Bigg] \Bigg\}^{1/2}Q_0 , \label{13} \\
\kappa &=& \tan^{-1}\bigg[ -\bigg( \f{\dot{f}(t)}{2\omega} + f(t) \tan \tilde{\theta}(t) \bigg) \bigg].
 \label{14}
\ea
We can easily check that the differentiation of $A_0$ with respect to time results in zero.
This means that $A_0$ is a time-constant.
Note that $A_0$ can also be expressed in a simple form as
\be
A_0 = \sqrt{\f{\epsilon\omega}{2\hbar}f(t_1)} Q_0, \label{15}
\ee
where $t_1$ is a particular time that is given by $t_1 = t_0 + (\pi/2-\theta_0)/\omega$.
From a direct differentiation of Eq. (\ref{14}) with time, we have
\be
\f{d\kappa}{dt} = - \f{\omega}{f(t)}. \label{16}
\ee
Hence, we can put $\kappa$ in the form
\be
\kappa = -\bigg(\omega \int_{t_0}^t f^{-1}(t') dt' + \theta\bigg), \label{17}
\ee
where $\theta$ is a phase.
Thus, we conclude that
\be
A(t) = A(t_0) e^{-i\omega T(t)}, \label{18}
\ee
where $A(t_0) = A_0 e^{-i\theta}$ and $T(t) = \int_{t_0}^t f^{-1}(t')dt'$.

\begin{figure}
\centering
\includegraphics[keepaspectratio=true]{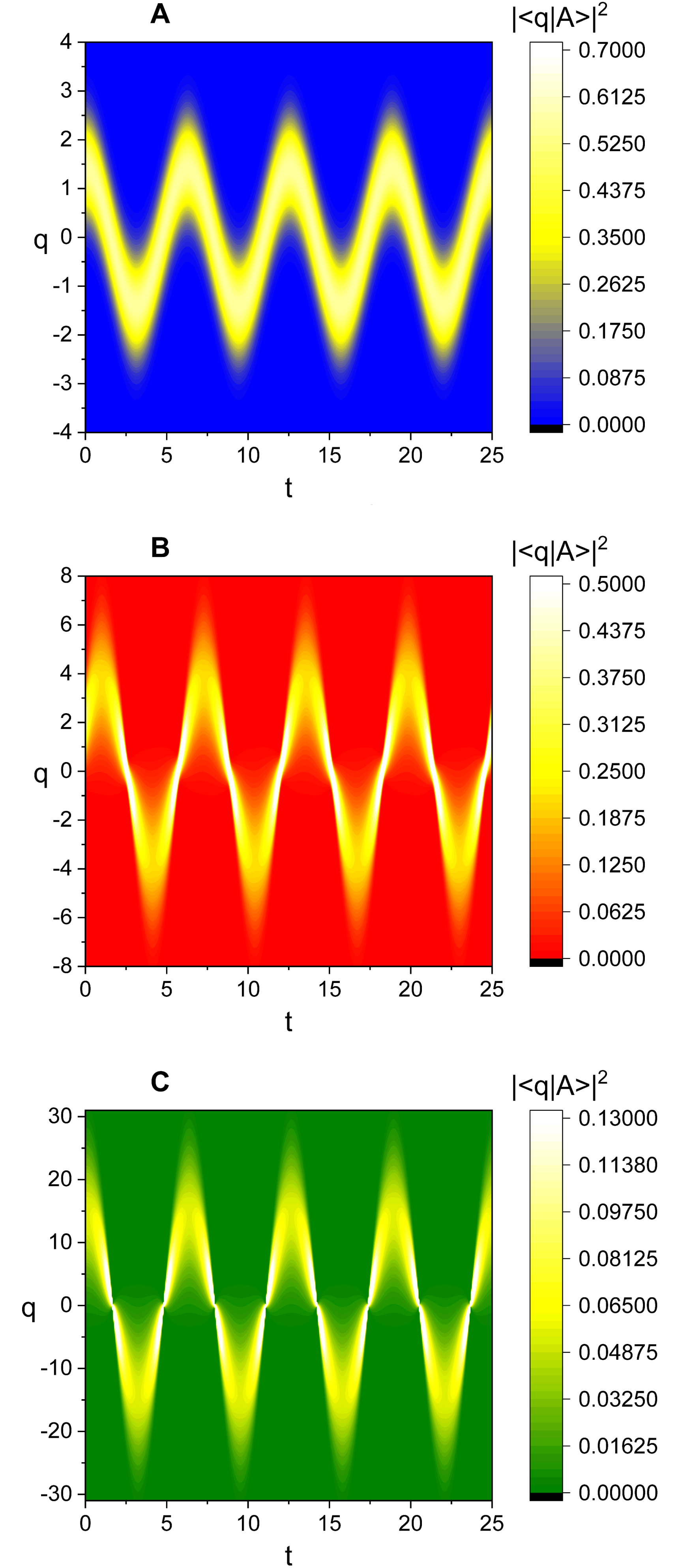}
\caption{\label{Fig1} Time evolution of the probability density $|\langle q |A \rangle|^2$
where ($c_1$, $c_2$) are ($1$, $1$) for A, ($5$, $2$) for B, and ($1$, $100$) for C.
The measure of nonstaticity is 0.00 for A, 2.37 for B, and 35.70 for C.
We have used $\omega=1$, $A_0=1$, $\hbar=1$, $\epsilon=1$, $t_0=0$, and $\varphi=\theta=0$.
We see from Eq. (\ref{5}) that the allowed values of $c_3$,
when $c_1$ and $c_2$ have been determined, are two: One is positive and the other is negative.
Among them, we choose a positive one as the value of $c_3$ in this and all subsequent figures for convenience.
 }
\end{figure}

The integration in $T(t)$ can be performed and this leads to \cite{nwh,gow}
\be
T(t) = G(t) - G(t_0) +{\mathcal G}(t)/\omega, \label{28} \\
\ee
where
\be
G(\tau) = \f{1}{\omega}\tan^{-1} \{ c_3+c_1\tan[\omega (\tau-t_0) +\varphi] \}, \label{29}
\ee
while ${\mathcal G}(t)=\pi \sum_{m=0}^{\infty}u[t-t_0-(2m+1)\pi/(2\omega)+\varphi/\omega]$ and
$u[x]$ is Heaviside step function.
By inserting Eq. (\ref{28}) with Eq. (\ref{29}) into Eq. (\ref{17}),
we can have the formula of $\kappa$.
Notice that $\kappa$ obtained in such a way is continuous over time because the discontinuity
originated from the characteristic of the arctangent function is
compensated by introducing the step function.

On the other hand, the former expression of $\kappa$ given in Eq. (\ref{14})
is discontinuous over time, although its time derivative gives the same result as that of
the new $\kappa$ which was mentioned a little while ago.
For this reason, we will use Eq. (\ref{17}) with  Eqs. (\ref{28}) and (\ref{29})
as the expression of $\kappa$ in the subsequent analysis of the nonstatic coherent wave.

\begin{figure}
\centering
\includegraphics[keepaspectratio=true]{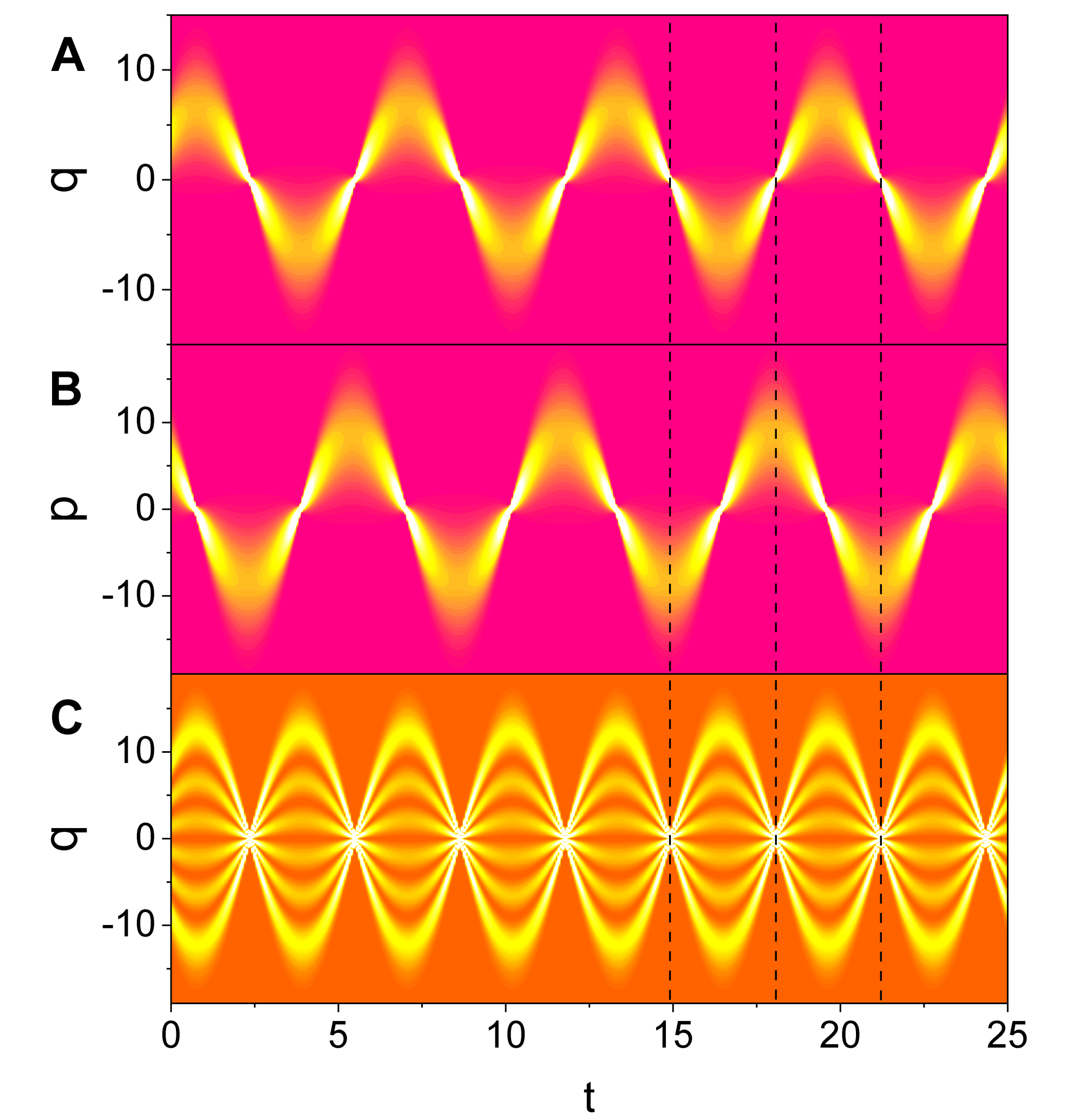}
\caption{\label{Fig2} Comparison between time evolutions of different probability densities for nonstatic waves.
A is $|\langle q |A \rangle|^2$, B is $|\langle p |A \rangle|^2$, and C is $|\langle q |\Psi_n \rangle|^2$
where $n=5$.
We have used ($c_1$, $c_2$)$=$($10$, $10$), $\omega=1$, $A_0=1$, $\hbar=1$, $\epsilon=1$, $t_0=0$, and
$\varphi=\theta=0$.
 }
\end{figure}

Because we know the complete formula of $A(t)$ from Eq. (\ref{18}) and subsequent equations
at this stage,
it is possible to investigate the characteristics of the wave function  given in Eq. (\ref{19}).
Figure 1 is the time evolution of the probability density $|\langle q |A \rangle|^2$
associated with that wave function.
The formula of $A(t)$ given in Eq. (\ref{18}) will also be used subsequently in order to
investigate other physical quantities in the coherent state.
By the way, if we take $c_3 =0$ and a different formula of $A$
similar to Eq. (\ref{9}) instead of the one given in Eq. (\ref{18}), the coherent state
developed here reduces to that of Ref. \cite{pos}.

$p$-space analysis of the coherent state may also be necessary
for the complete understanding of the evolution of the nonstatic wave.
$p$-space eigenfunction of the invariant operator, i.e., the wave function in $p$-space
can be obtained from the Fourier transformation:
\be
\langle {p}|A \rangle  = \frac{1}{\sqrt{2\pi \hbar}}
\int_{-\infty}^{\infty} \langle q|A \rangle
e^{-i {p}{q}/\hbar} d q.  \label{20}
\ee
The exact formula of $\langle {p}|A \rangle$ is represented in {\bf METHODS} section.

The three kinds of the probability densities, $|\langle {q}|A \rangle|^2$, $|\langle {p}|A \rangle|^2$,
and $|\langle {q}|\Psi_n \rangle|^2$ ($n=1,2,3,\cdots$),
have been compared to each other in Fig. 2, where
$\langle {q}|\Psi_n \rangle$ are Fock-state wave functions that were previously
investigated in Ref. \cite{nwh}. The formula of $\langle {q}|\Psi_n \rangle$ has been
represented in {\bf METHODS} section for convenience.
The phase difference between the evolutions of $|\langle {q}|A \rangle|^2$ and $|\langle {p}|A \rangle|^2$
is $\pi/2$.
From a careful comparison of Fig. 2(A) with Fig. 2(C), we confirm that
$|\langle {q}|A \rangle|^2$ constitutes a node (a belly) whenever $|\langle {q}|\Psi_n \rangle|^2$
a node (a belly).
From this, we can confirm the similarity between the nonstatic evolutions of
the coherent-state wave and the Fock-state wave.
\\
\\
{\bf Quantum Energy and Quadrature Fluctuations} \\
As the wave becomes nonstatic, the evolutions of related physical quantities may also
deviate from their standard patterns.
To see this in a quantitative way, lets consider quantum energy and quadrature fluctuations for instance.

\begin{figure}
\centering
\includegraphics[keepaspectratio=true]{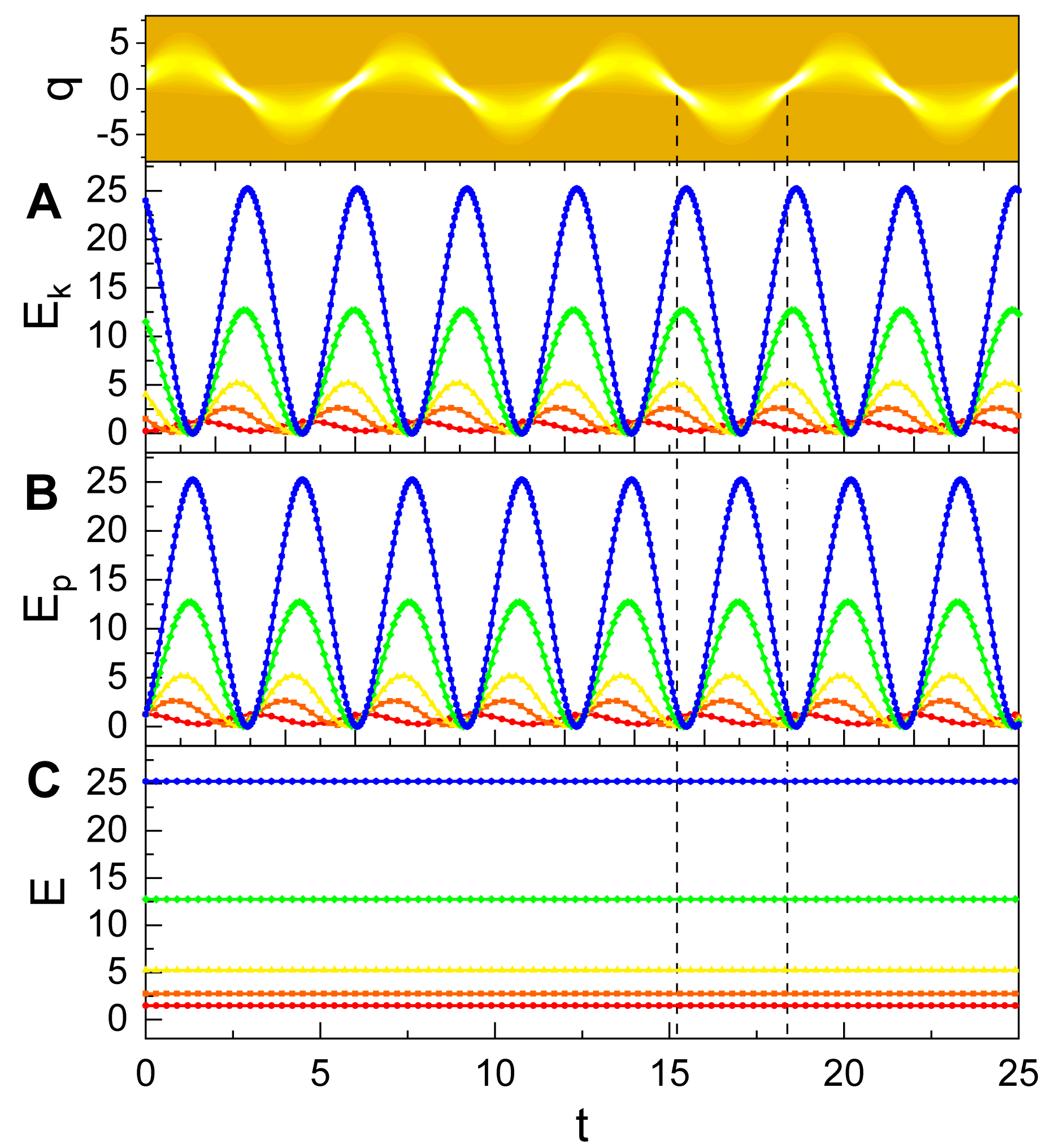}
\caption{\label{Fig3} Time evolution of quantum electric energy $E_{\rm k}$,
quantum magnetic energy $E_{\rm p}$,
and total quantum energy $E$ for several different values of $c_1$.
The value of $c_1$ that we have chosen is 1 for red, 2 for orange, 4 for yellow, 10 for green,
and 20 for blue lines.
We have used $c_2=1$, $\omega=1$, $A_0=1$, $\hbar=1$, $\epsilon=1$, $t_0=0$, and
$\varphi=\theta=0$. The evolution of the probability density for the case of $c_1=4$
is provided in upper part: this is associated to the quantities of yellow lines.
 }
\end{figure}

\begin{figure}
\centering
\includegraphics[keepaspectratio=true]{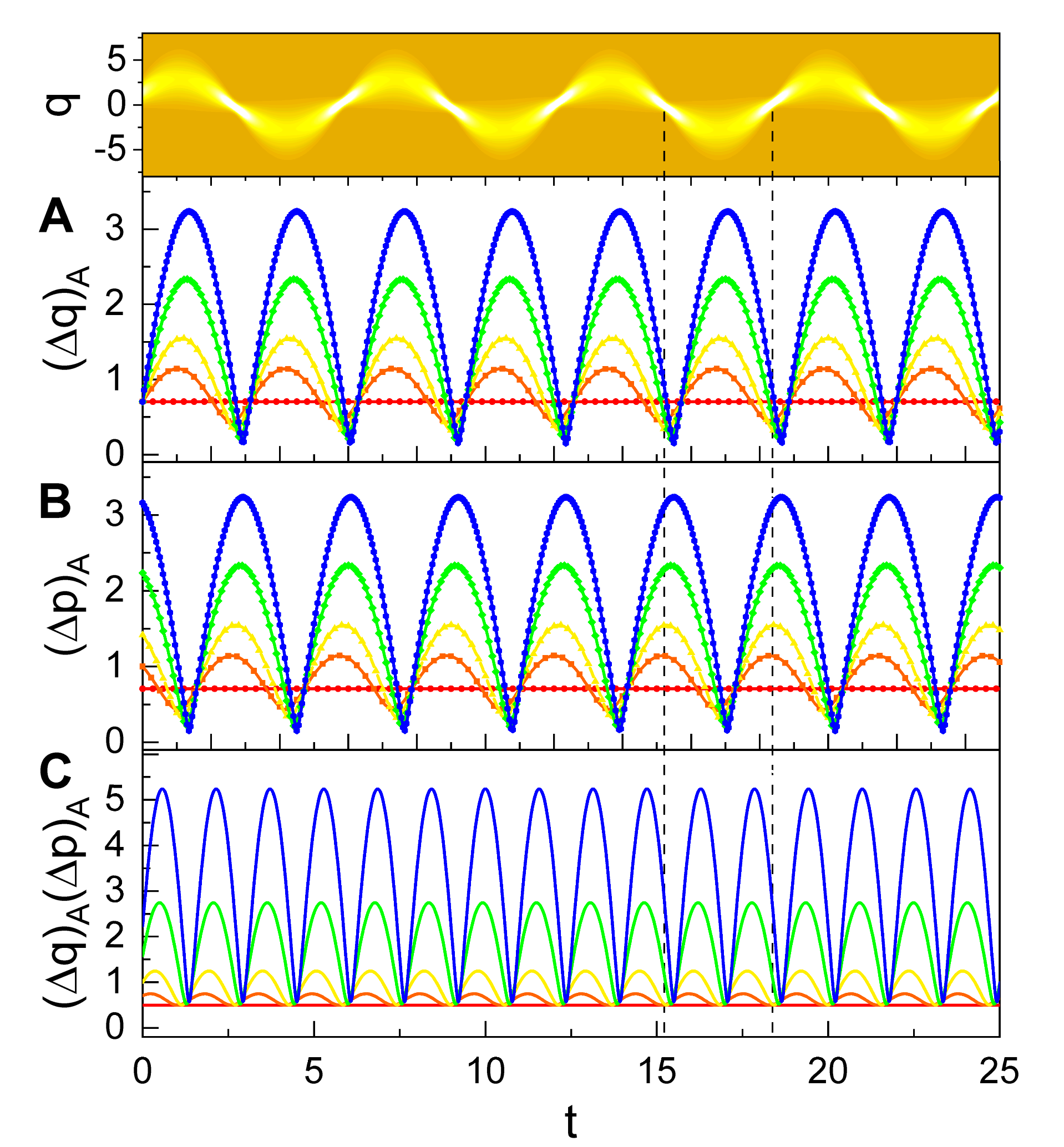}
\caption{\label{Fig4} Time evolution of quadrature
fluctuations $(\Delta q)_A$ (A), $(\Delta p)_A$ (B), and the uncertainty product
$(\Delta q)_A(\Delta p)_A$ (C) for several different values of $c_1$.
The value of $c_1$ that we have chosen is 1 for red, 2 for orange, 4 for yellow, 10 for green,
and 20 for blue lines.
We have used $c_2=1$, $\omega=1$, $\hbar=1$, $\epsilon=1$, $t_0=0$, and $\varphi=0$.
In upper part, the evolution of the probability density is shown with the choice of $c_1=4$.
 }
\end{figure}

As shown previously, the Hamiltonian is composed of two terms
associated with electric energy ${\hat{p}^2}/{(2\epsilon)}$
and magnetic energy $\epsilon\omega^2 \hat{q}^2/2$.
From the evaluation of the expectation values of them in the coherent state
using Eqs. (\ref{6-1}) and (\ref{7-1}), we have the
formula of the electric energy and the magnetic energy
from quantum-mechanical point of view as
\ba
E_{\rm k} &=& -\f{\hbar\omega}{4f(t)} \Bigg[ \left(1+i \f{\dot{f}(t)}{2\omega}\right)^2 A^2
+ \left(1-i \f{\dot{f}(t)}{2\omega}\right)^2 A^{*2}  \nonumber \\
& &-\left(1+ \f{[\dot{f}(t)]^2}{4\omega^2}\right) (2A^*A+1) \Bigg], \label{sa26} \\
E_{\rm p} &=& \f{1}{4} \hbar\omega f(t) [A^2 + A^{*2} + 2A^*A +1]. \label{sa27}
\ea
The time behaviors of these are shown in Fig. 3.
Depending on the wave variation over time, both the electric
and the magnetic energies vary periodically. This can be regarded as the manifestation of
wave nonstaticity.
The electric energy is largest at nodes, whereas the magnetic energy
is largest at the bellies.
However, the total quantum energy does not vary over time and this consequence agrees
with the universal physical law of energy conservation.

The fluctuation of an
observable $\hat{O}$ in the coherent state can be defined as $(\Delta O)_A =
[\langle \hat{O}^2 \rangle - \langle \hat{O} \rangle^2]^{1/2}$
where $\langle \cdots \rangle = \langle A|\cdots |A\rangle$.
Using this, the fluctuations of canonical variables
represented in Eqs. (\ref{6-1}) and (\ref{7-1}) are obtained, such that
\ba
(\Delta q)_A &=& \Bigg(\f{\hbar f(t)}{2\epsilon\omega}\Bigg)^{1/2}, \label{sa28} \\
(\Delta p)_A &=& \Bigg[\f{\hbar  \epsilon\omega}{2 f(t)}
\Bigg( 1 + \f{[\dot{f}(t)]^2}{4\omega^2} \Bigg)\Bigg]^{1/2}. \label{sa29}
\ea
We also readily have the corresponding uncertainty product as
\be
(\Delta q)_A(\Delta p)_A = \f{\hbar}{2}
\Bigg( 1 + \f{[\dot{f}(t)]^2}{4\omega^2} \Bigg)^{1/2}. \label{sa30}
\ee
The time evolutions of the fluctuations $(\Delta q)_A$ and $(\Delta p)_A$
together with $(\Delta q)_A(\Delta p)_A$ are represented in Fig. 4.
These also exhibit periodic behaviors over time.
$(\Delta q)_A$ is largest at bellies, while $(\Delta p)_A$ is largest at nodes.
On the other hand, $(\Delta q)_A(\Delta p)_A$ is smallest at both nodes and bellies.
By comparing Fig. 4(A) with
Fig. 4(B), we see that the variation patterns of $(\Delta q)_A$ and $(\Delta p)_A$ are
the same as each other except for the difference in the phase between them.
\\
\\
{\bf Squeezing Effects and Nonclassicality} \\
From Fig. 4(A,B), we see that both the uncertainties $(\Delta q)_A$ and $(\Delta p)_A$ can be lowered below their
standard quantum levels.
If $(\Delta q)_A$ is lower than its standard quantum level, $(\Delta p)_A$ is larger than
its standard quantum one and vice versa.
This means that the nonstatic coherent state resembles the squeezed state on one hand.
In order to see in more detail about this, let us consider the Wigner distribution function
which is defined in the form 
\be
W (q,p,t) = \f{1}{\pi \hbar} \int_{-\infty}^\infty
\langle A |q+y \rangle \langle q-y |A \rangle
e^{2ipy/\hbar} dy. \label{sa31}
\ee
\begin{figure}
\centering
\includegraphics[keepaspectratio=true]{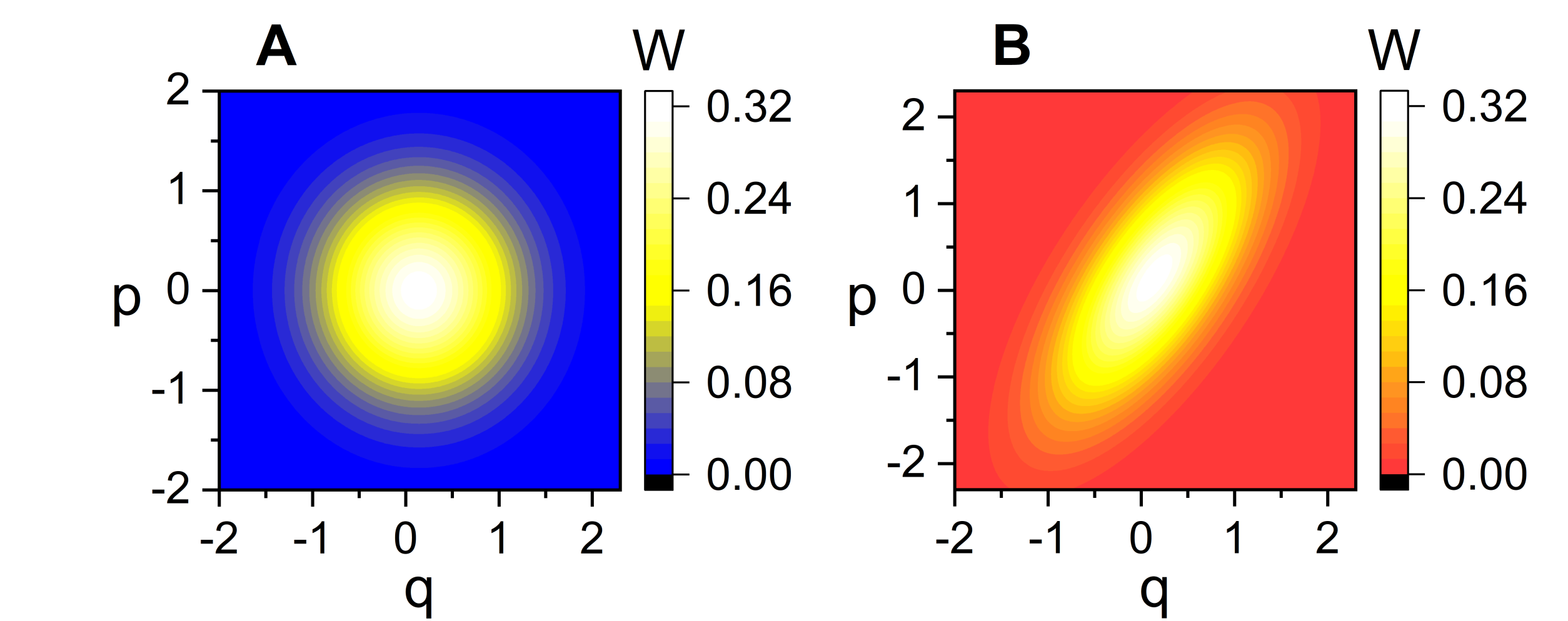}
\caption{\label{Fig5} Comparison of the density plot of $W$ for static (A) and nonstatic (B) light waves at $t=0$
where ($c_1$, $c_2$) are chosen as ($1$, $1$) for A and ($2$, $1$) for B.
We have used $\omega=1$, $A_0=0.1$, $\hbar=1$, $\epsilon=1$, $t_0=0$, and $\varphi=\theta=0$.
 }
\end{figure}
A straightforward evaluation of this using Eq. (\ref{19}) results in
\ba
W (q,p,t) &=& \f{1}{\pi \hbar} \exp\Bigg[-\zeta(t)q^2 - \Bigg( \f{\sqrt{\zeta(t)}}{2\omega}\dot{f}(t)q
- \f{p}{\sqrt{\zeta(t)}\hbar} \Bigg)^2 \nonumber \\
& &+ \sqrt{2\zeta(t)}\bigg( (A+A^*)+i \f{\dot{f}(t)}{2\omega}(A-A^*) \bigg)q
- \f{i}{\hbar}\sqrt{\f{2}{\zeta(t)}}(A-A^*)p -2|A|^2\Bigg]. \label{sa32}
\ea
The graphical illustration for this outcome is given in Fig. 5.
By comparing Fig. 5(B) with Fig. 5(A), we see that $W$
is squeezed in a certain direction in phase space as the state becomes nonstatic.
The time evolution of the Wigner distribution function for a nonstatic wave with a relatively
large amplitude ($A_0$)
is shown in Fig. 6.
\begin{figure}
\centering
\includegraphics[keepaspectratio=true]{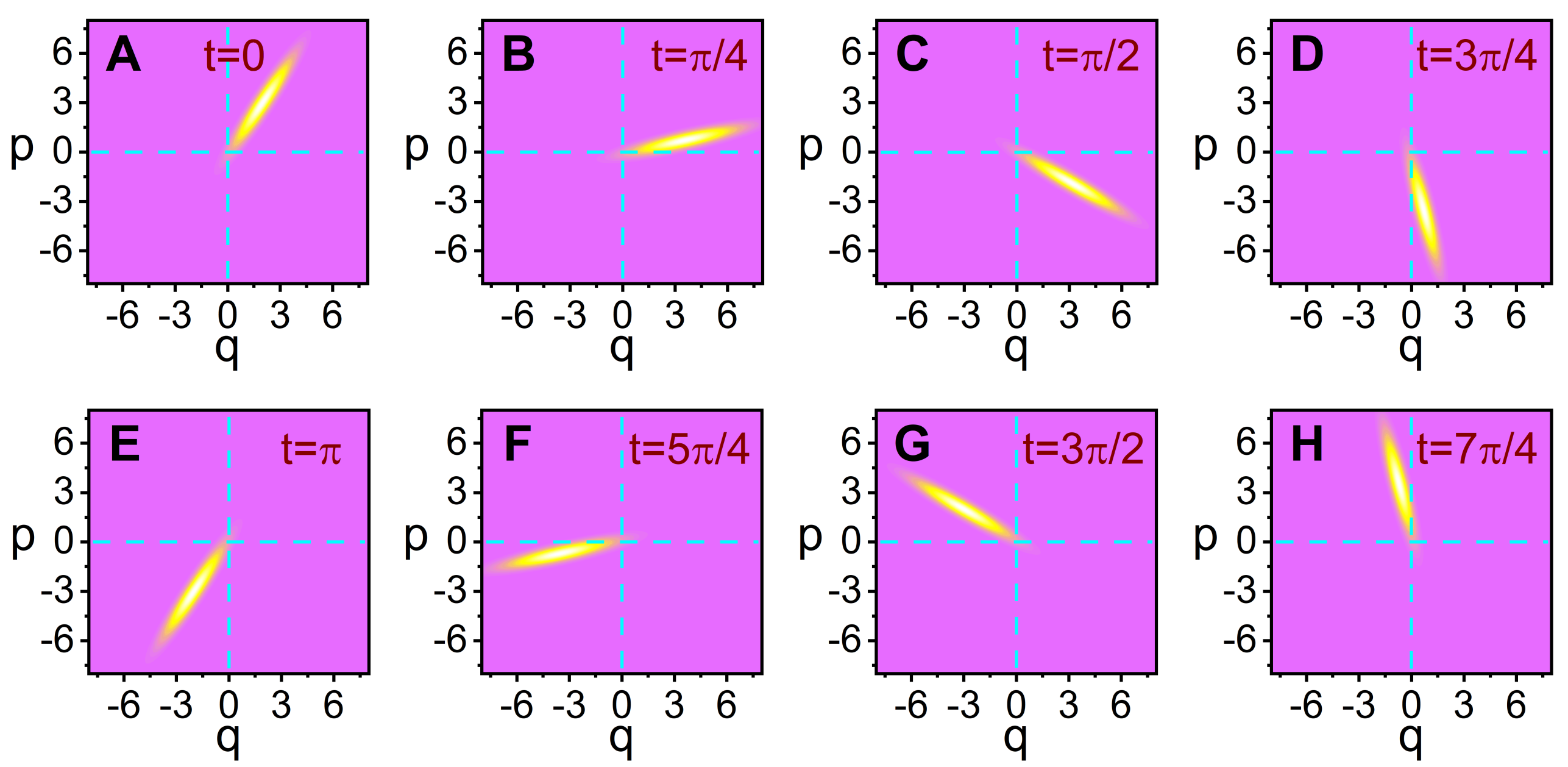}
\caption{\label{Fig6} Density plots for the time evolution of $W$
where the considered time is represented in each panel.
The nonstaticity parameters chosen here are ($c_1$, $c_2$)$=$($5$, $2$);
this choice corresponds to that of Fig. 1(B).
We have chosen the amplitude as $A_0=1$, which is
very large compared to that in Fig. 5.
All other parameters taken here are the same as those of Fig. 5.
 }
\end{figure}
From this figure, we see that the bar that represents the
squeezed distribution contour rotates as time goes by.
Not only the center of the bar rotates clockwise with respect to the
origin of coordinates, but the bar itself rotates clockwise about its center as well.
The periods of both kinds of rotations are the same as each other and they are
given by $t_{\rm p} = 2\pi/\omega$.
Whereas the former kind of rotation is usual, the latter kind is clearly
related to the fundamental light-wave nonstaticity,
such as the time variation of the uncertainties and the appearance of bellies and nodes in
the wave evolution.
Let us look into Fig, 6
in connection with Fig. 1(B) that was taken the same values of $c_1$ and $c_2$.
We confirm that the values of
$(\Delta q)_A$ in panels B and F in Fig. 6 are relatively high,
whereas they correspond to the instants of time at which the displacements in Fig. 1(B) is nearly
highest.
On the other hand, they are small in panels D and H, whereas these cases
correspond to the instants where the wave forms nearly a node in Fig. 1(B).
This outcome agrees with the result of Fig. 4(A) which exhibits
that $(\Delta q)_A$ is small around a node and high around a belly.
From this analysis, we can understand the mechanism of squeezing that arises in the nonstatic coherent state.
Squeezing effects in a quantum state is a well-known nonclassical property.

We now analyze the nonclassicality of the nonstatic state in more detail in relation to
the standard description of light waves.
To this end, we introduce the usual annihilation operator:
\be
\hat{a}= \sqrt{\f{\epsilon\omega}{2\hbar }} \hat{q} + i\f{\hat{p}}{\sqrt{2\epsilon\omega\hbar}} ,  \label{sa33}
\ee
and its Hermitian adjoint $\hat{a}^\dagger$ that is the creation operator.
Then, Eq. (\ref{7}) can be rewritten in terms of them to be
\be
\hat{A} = \mu(t) \hat{a} + \nu(t) \hat{a}^\dagger, \label{sa36}
\ee
where
\ba
\mu(t) &=& \f{1}{2\sqrt{f(t)}}\left(1- i \f{\dot{f}(t)}{2\omega}\right) + \f{\sqrt{f(t)}}{2}, \label{sa37} \\
\nu(t) &=& \f{1}{2\sqrt{f(t)}}\left(1- i \f{\dot{f}(t)}{2\omega}\right) - \f{\sqrt{f(t)}}{2}. \label{sa38}
\ea
Notably, $\mu(t)$ and $\nu(t)$ satisfy the relation
\be
|\mu|^2 - |\nu|^2 =1. \label{sa39}
\ee
Based on the above expressions, we can also write Eqs. (\ref{sa28}) and (\ref{sa29}) as
\ba
(\Delta q)_A &=& \bigg(\f{\hbar }{2\epsilon\omega}\bigg)^{1/2} (\mu -\nu), \label{sa40} \\
(\Delta p)_A &=& \bigg(\f{\hbar  \epsilon\omega}{2}\bigg)^{1/2} |\mu +\nu|. \label{sa41}
\ea
If $\mu=1$ and $\nu=0$ (or $c_1=c_2=1$), these reduce to
standard uncertainties in each quadrature, which are represented with
a red curve in panels A and B in Fig. 4, respectively.
Thus, we confirm that the time variation of the uncertainties shown in Fig. 4 is determined depending purely on
the time variations of $\mu$ and $\nu$ that follow Eqs. (\ref{sa37}) and  (\ref{sa38}).

\begin{figure}
\centering
\includegraphics[keepaspectratio=true]{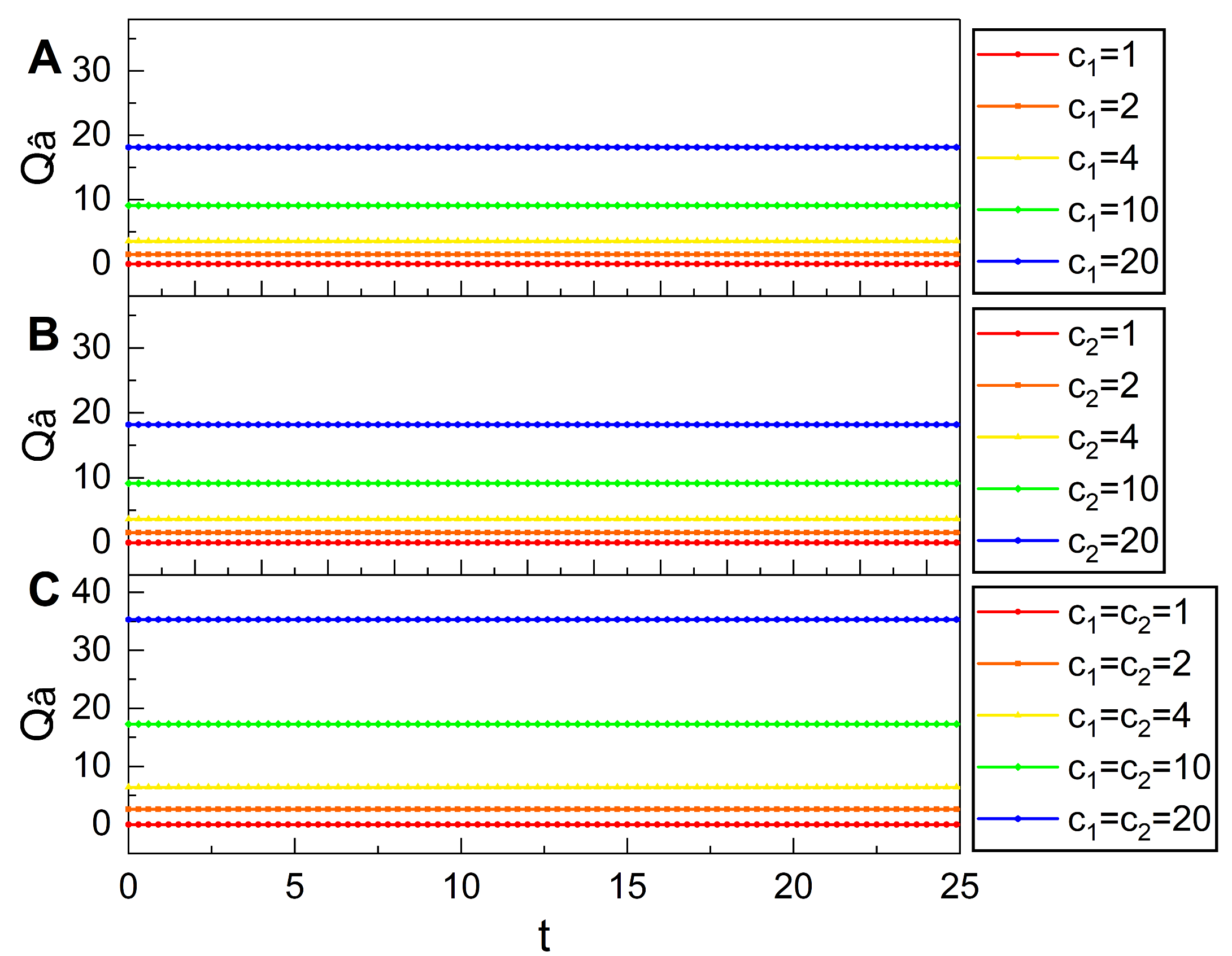}
\caption{\label{Fig7} Mandel's
parameter $Q_{\hat a}$ versus $t$ for several different values of $c_1$ (A), $c_2$ (B), and $c_1$ and $c_2$ (C).
We have used $c_2=1$ for A, $c_1=1$ for B, $\omega=1$, $A_0=1$, $t_0=0$, and $\varphi=\theta=0$.
 }
\end{figure}

As a measure of nonclassicality, we see Mandel's Q parameter \cite{mand} for this system.
$\hat{a}$ and $\hat{a}^\dagger$ can be used for estimating nonclassicality of the nonstatic state
relative to the standard description of light waves because standard quantum states are
described in terms of them.
Considering this, it is possible to represent the Q parameter as \cite{mand}
\be
Q_{\hat a} = \f{[(\Delta(\hat{a}^\dagger \hat{a}))_A]^2
- \langle \hat{a}^\dagger \hat{a} \rangle}{\langle \hat{a}^\dagger \hat{a} \rangle}, \label{sa44}
\ee
where the definition of $\langle \cdots \rangle$ is still the same as the previous one which is
$\langle A|\cdots |A\rangle$. While there is no upper bound for the Q parameter,
its minimal value allowed in quantum mechanics is $-1$.
If $-1 \leq Q_{\hat a} < 0$, the field follows sub-Poissonian statistics, whereas it follows
super-Poissonian statistics when $Q_{\hat a} > 0$.
The light-wave description, on the other hand, reduces to Poissonian statistics
in the case $Q_{\hat a} = 0$.

To evaluate the Q parameter, let us represent $\hat{a}$ and $\hat{a}^\dagger$ in the form
\ba
\hat{a} &=& \mu^* \hat{A} -\nu \hat{A}^\dagger, \label{sa42}  \\
\hat{a}^\dagger &=& \mu \hat{A}^\dagger -\nu^* \hat{A}. \label{sa43}
\ea
If we insert these two formulae into Eq. (\ref{sa44}), the problem is treated in terms
of some arrays of $\hat{A}$ and $\hat{A}^\dagger$ instead of the standard ladder operators.
We then rearrange the elements of Eq. (\ref{sa44}) in a way that each array
of ladder operators in their representation being the normal order \cite{fpa}.
Thus, with the help of Eq. (\ref{8}), we eventually have
\ba
\langle \hat{a}^\dagger \hat{a}\rangle &=& (|\mu|^2+|\nu|^2) |A|^2
-\mu\nu A^{*2}-\mu^*\nu^* A^{2}+|\nu|^2, \label{sa45} \\
{[(\Delta(\hat{a}^\dagger \hat{a}))_A]}^2 &=& (|\mu|^4+6|\mu|^2|\nu|^2+|\nu|^4)|A|^2
-2 (|\mu|^2+|\nu|^2)(\mu\nu A^{*2}+\mu^*\nu^* A^{2})  \nonumber \\
& &+2|\mu|^2 |\nu|^2. \label{sa46}
\ea
The dependence of Q parameters on $c_1$ and $c_2$ is illustrated in Fig. 7.
From Fig. 7(A) (Fig. 7(B)), we confirm that $Q_{\hat a}$ for nonstatic states is larger than 0
and it increases as $c_1$ ($c_2$) grows.
Figure 7(C) shows that the value of the Q parameter is much higher when both $c_1$ and $c_2$ are large.
Interestingly, the Q parameter does not vary over time although it is represented in terms of
the sinusoidal-like time function $f(t)$.
There are also many systems in which the Q parameter depends on time \cite{tdq1,tdq2,tdq3}.
The Q parameter reduces
to 0 in the standard-coherent-state limit ($c_1=c_2=1$) as expected.

Since $Q_{\hat a} > 0$ unless $c_1=c_2=1$, the nonstatic state considered here is described by
super-Poissonian statistics. Other systems in which the photon distribution is governed by
super-Poissonian statistics are found in Refs. \cite{sps1,sps2,sps3,sps4}.
While it is possible to attain photon bunching via the super-Poissonian distribution of photons,
the noise in the associated photo-count is higher than the one for the standard coherent state \cite{fpa}.
Strong electromagnetic fields with enhanced photon bunching is important
for controlling multiexciton processes in core-shell nanocrystals \cite{sps1}
and intensity correlations in EIT (Electromagnetically Induced Transparency) media \cite{sps2}.
\\
\\
{\bf CONCLUSION
\vspace{0.2cm}}
\\
The quantum mechanical behavior of a nonstatic light wave in the coherent state has been analyzed.
A generalized annihilation operator was introduced and its eigenfunction
which plays the wave function of light was derived.
We confirmed that the modification in the evolution pattern of the probability density
reflects the details of the wave nonstaticity.
The departure of the periodical wave evolution from that of the well-known ordinary wave
becomes distinct as the degree of nonstaticity increases.

The amplitude of the wave collapses and expands in turn as a manifestation of its nonstaticity
like the behavior of the Fock-state nonstatic wave.
A node takes place in the graphic of wave evolution in quadrature space
whenever the wave passes through $q=0$, whereas
a belly takes place whenever the displacement of the wave is instantaneously largest.
In fact, the instants of time where a node (or a belly) occurs are the same as
those in the Fock states.
The wave in conjugate $p$-quadrature space also exhibits similar pattern of nonstaticity,
but the phase of its evolution precedes $\pi/2$ compared to that of the $q$-space wave.

The electric and magnetic energies of the wave also vary according to
the characteristics of nonstaticity in the evolution of the wave.
The electric energy
is largest at the nodes, whereas the magnetic energy is largest at the bellies.
However, the total wave energy does not vary over time,
leading energy being conserved even if the wave exhibits nonstaticity.
The fluctuations of quadratures $q$ and $p$ also exhibit periodic behaviors
due to nonstaticity of the wave.
$(\Delta q)_A$ is largest at bellies and $(\Delta p)_A$ is largest at nodes, while
the corresponding uncertainty product $(\Delta q)_A (\Delta p)_A$ is in contrast smallest at
both bellies and nodes.

We have confirmed that the above characteristics of wave nonstaticity in the coherent state
can be explained by means of the analysis of the Wigner distribution function.
As the wave becomes nonstatic, the contour of the Wigner distribution function in its phase-space plot
exhibits squeezing and rotates clockwise with respect to
its center. This rotation is responsible for various effects of the wave nonstaticity.
We also confirmed that the Mandel's Q parameter for the nonstatic wave is larger than unity.
From this, the wave follows super-Poissonian statistics.
\\
\\
{\bf METHODS  \vspace{0.2cm}} \\
{\bf Methods Summary} \\
To describe the nonstaticity of a light wave, we introduce an invariant operator that obeys
the Liouville von-Neumann equation.
The invariant operator is expressed in terms of generalized annihilation and
creation operators ($\hat{A}$ and $\hat{A}^\dagger$).
By solving the eigenvalue equation of $\hat{A}$, we establish a coherent state
that exhibits the characteristic of nonstaticity.
Based on the wave function in this state, we investigate light wave nonstaticity.
The quantum energy, quadrature fluctuations, the Wigner distribution function, and
Mandel's Q parameter in the nonstatic coherent state are
derived using the wave function.
\\
\\
{\bf Wave Functions in the Fock States} \\
Fock-state wave functions with nonstaticity are given by \cite{nwh}
\be
\langle q |\Psi_n(t) \rangle = \langle q |\Phi_n(t) \rangle \exp [i\gamma_n(t) ], \label{M32}
\ee
where $\langle q |\Phi_n(t) \rangle$ are eigenfunctions of $\hat{I}$
(given in Eq. (\ref{2}) or in Eq. (\ref{6})) and $\gamma_n (t)$ wave phases, of which
formulae are of the form
\ba
\langle q |\Phi_n(t) \rangle &=& \left({\f{\zeta(t)}{\pi}}\right)^{1/4} \f{1}{\sqrt{2^n
n!}} H_n \left( \sqrt{\zeta(t)} q \right) \exp \left[
- \f{1}{2}\zeta(t) \left(1-i\f{\dot{f}(t)}{2\omega}\right) q^2 \right], \label{M33} \\
\gamma_n(t) &=& {-\omega (n+1/2) \int_{t_0}^t f^{-1} (t') dt'} + \gamma_n (t_0).
 \label{M34}
\ea
while $H_{n}$ are $n$th order Hermite polynomials.
\\
\\
{\bf The Eigenfunction in $p$-quadrature Space} \\
The eigenfunction in $p$-quadrature is easily evaluated from Eq. (\ref{20}) and it results in
\ba
\langle {p}|A \rangle &=& \frac{1}{\sqrt[4]{\pi \zeta }}
\frac{1}{\sqrt{\hbar[1-i\dot{f}(t)/(2\omega)]}}
\exp\Bigg[ -\f{p^2+2i\sqrt{2\zeta}A\hbar p}{2\zeta \hbar^2[1-i\dot{f}(t)/(2\omega)]} \nonumber \\
& &+ \f{1+i\dot{f}(t)/(2\omega)}{2[1-i\dot{f}(t)/(2\omega)]}A^2 -\f{1}{2}|A|^2 \Bigg].  \label{20-}
\ea


\end{document}